\documentclass[prc,aps,twocolumn,showpacs,amssymb,superscriptaddress]{revtex4}

\usepackage{amsmath}
\usepackage{graphicx}
\usepackage{dcolumn}
\usepackage{float}
\begin{document}

\title{Shell-model study of quadrupole collectivity in light tin isotopes}

\author{L. Coraggio}
\affiliation{Istituto Nazionale di Fisica Nucleare, \\
Complesso Universitario di Monte  S. Angelo, Via Cintia - I-80126 Napoli,
Italy}
\author{A. Covello}
\affiliation{Dipartimento di Fisica, Universit\`a
di Napoli Federico II, \\
Complesso Universitario di Monte  S. Angelo, Via Cintia - I-80126 Napoli,
Italy}
\author{A. Gargano}
\affiliation{Istituto Nazionale di Fisica Nucleare, \\
Complesso Universitario di Monte  S. Angelo, Via Cintia - I-80126 Napoli,
Italy}
\author{N. Itaco}
\affiliation{Istituto Nazionale di Fisica Nucleare, \\
Complesso Universitario di Monte  S. Angelo, Via Cintia - I-80126 Napoli,
Italy}
\affiliation{Dipartimento di Fisica, Universit\`a
di Napoli Federico II, \\
Complesso Universitario di Monte  S. Angelo, Via Cintia - I-80126 Napoli,
Italy}
\author{T. T. S. Kuo}
\affiliation{Department of Physics, SUNY, Stony Brook, New York 11794}
\date{\today}

\begin{abstract}
A realistic shell-model study is performed for neutron-deficient tin
isotopes up to mass $A=108$.
All shell-model ingredients, namely two-body matrix elements,
single-particle energies, and effective charges for electric quadrupole
transition operators, have been calculated by way of the many-body
perturbation theory, starting from a low-momentum interaction derived
from the high-precision CD-Bonn free nucleon-nucleon potential.
The focus has been put on the enhanced quadrupole collectivity of
these nuclei, which is testified by the observed large $B(E2;0_1^+
\rightarrow 2^+_1)$s.
Our results evidence the crucial role played by the $Z=50$ cross-shell
excitations that need to be taken into account explicitly to obtain a
satisfactory theoretical description of light tin isotopes.
We find also that a relevant contribution comes from the calculated
neutron effective charges, whose magnitudes exceed the standard
empirical values.
An original double-step procedure has been introduced to reduce
effectively the model space in order to overcome the computational
problem.
\end{abstract}


\maketitle

Light tin isotopes have been an interesting laboratory since the early
90s, when the experimental efforts toward the observation of the
doubly-closed $^{100}$Sn became a sort of search of the ``Holy Grail''.
This has led to a certain amount of data that have improved our
understanding of the structure of neutron-deficient isotopes,
providing also a challenging ground for shell-model calculations.
As a matter of fact, the study of light tin isotopes opened the way to
a new generation of realistic shell-model calculations
\cite{Engeland93,Hjorth95,Andreozzi96a}, an approach that has then
flourished in the last two decades.

In the last few years, a renewed experimental interest has arised in
studying these nuclei, especially with the help of intermediate-energy
Coulomb-excitation experiments that are able to provide information on
the electric quadrupole-excitation properties.

In particular, in 2013 two papers reported about the measurement of
the electric-quadrupole $0^+_1 \rightarrow 2^+_1$ transition rate in
$^{104}$Sn.
The first work was performed at GSI \cite{Guastalla13}, and the
measured value is $B(E2;0^+_1 \rightarrow 2^+_1)=1000 \pm 400~e^2{\rm
  fm}^4$.
This value fits well with the predictions of realistic shell-model
calculations \cite{Banu05}, where empirical effective charges have
been employed, both when the model space is made up by only
neutron orbitals above the $^{100}$Sn core and when also proton
excitations coming from the proton $0g_{9/2}$ orbital are included
considering $^{90}$Zr as a closed core.

The second paper reported the results of an experiment carried out at
the National Superconducting Cyclotron Laboratory (NSCL) at the
Michigan University \cite{Bader13}.
In this work the measured value of the $^{104}$Sn $B(E2;0^+_1 \rightarrow
2^+_1)$ is larger, $1800 \pm 370~e^2{\rm fm}^4$, and
disagrees more than one sigma with the value of Ref. \cite{Guastalla13}.

Very recently, another measurement of this transition probability has
been performed at RIKEN \cite{Doornenbal14}, wherein the result has
been obtained from absolute Coulomb excitation cross sections.
The reported value of the $^{104}$Sn $B(E2;0^+_1 \rightarrow
2^+_1)$ is $1730 \pm 280~e^2{\rm fm}^4$, consistent with the result of
Ref. \cite{Bader13}.

The value obtained in the last two papers is quite large and is not
reproduced by shell-model calculations, even when proton degrees of
freedom are explicitly taken into account.
In fact, various calculations 
\cite{Banu05,Vaman07,Guastalla13,Qi12,Back13,Bader13} have been recently
performed using $^{100}$Sn and $^{90}$Zr as closed cores, and they all
predict $B(E2)$ values too small for the neutron-deficient tin
isotopes as compared to the experimental ones.
Moreover, no significant improvement is obtained by including also
neutron excitations across the $N=50$ shell closure, these
excitations leading only to a slight increase of the  $B(E2)$s
\cite{Guastalla13}. 

This background has been the main motivation to perform realistic
shell-model calculations for neutron-deficient tin isotopes, using
both the standard $^{100}$Sn neutron-only model space and a larger one
that includes $Z=50$ cross-shell excitations with $^{88}$Sr as an
inert core.
The main new elements of these calculations with respect to the
previous ones are the inclusion of proton excitations from the
$1p_{1/2}$ orbital and the use of microscopic effective charges, as
well as an original procedure to reduce the large model space when
considering $^{88}$Sr as a closed core.

Another motivation is to revisit a region that, as
mentioned before, has been for us the starting point for the
investigation of the reliability of realistic shell-model
calculations, going from the $p$-shell region up to nuclei around
doubly-closed $^{208}$Pb core
\cite{Coraggio01,Coraggio10a,Coraggio07b,Coraggio14b,Coraggio09b,Coraggio99}.

In our shell-model calculation we start from the high-precision
nucleon-nucleon potential CD-Bonn \cite{Machleidt01b}, whose
high-momentum repulsive components are smoothed out using the $V_{\rm
  low-k}$ approach \cite{Bogner02} so as to derive an effective
hamiltonian $H_{\rm eff}$ by way of the time-dependent perturbation
theory \cite{Kuo71,Coraggio09a}.
The chosen cutoff momentum is $\Lambda=2.6$ fm$^{-1}$, and $H_{\rm
  eff}$ has been calculated including diagrams up to third-order in
$V_{\rm low-k}$.

From the effective hamiltonian both single-particle (SP) energies and
two-body matrix elements (TBME) of the residual interaction have been
obtained \cite{Coraggio12a}, and we have derived consistently the
effective charges of the electric quadrupole operators at the same
perturbative order. 

\begin{figure}[t]
\begin{center}
\includegraphics[scale=0.38,angle=0]{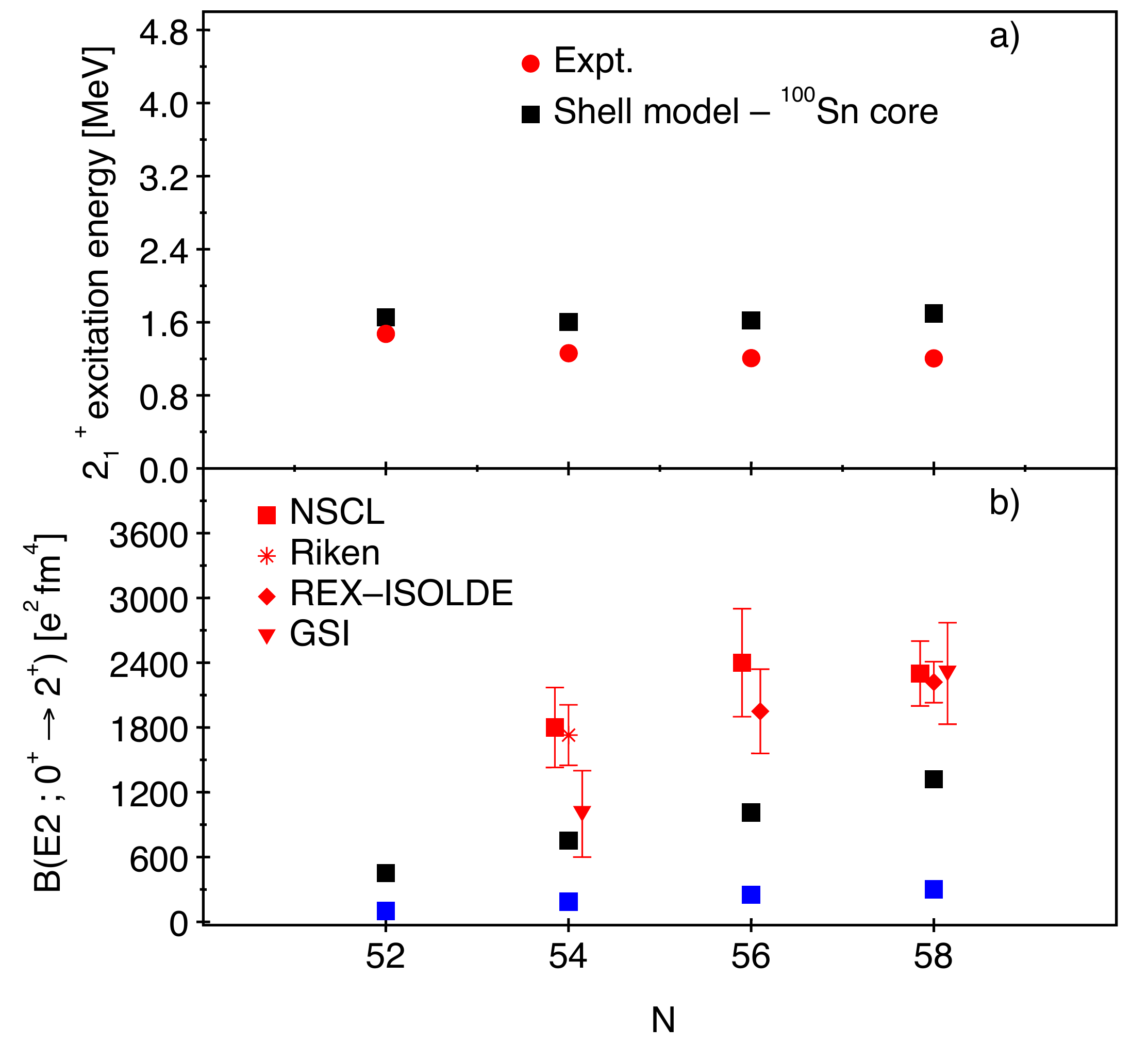}
\caption{(Color online) (a) Experimental
  \cite{Banu05,Vaman07,Ekstrom08,Guastalla13,Bader13,Doornenbal14}
  (red symbols) and calculated (black squares) excitation energies of the
  yrast $J^{\pi}=2^+$ states and (b) $B(E2;2^+_1 \rightarrow 0^+_1)$
  transition rates for tin isotopes up to $N=58$, when using neutron
  effective charges (I) reported in Table \ref{effchn}. Blue squares refer to
  calculations with $e^{\rm emp}_n =0.5e$. Shell model calculations
  have been performed using only neutron degrees of freedom (see text
  for details).} 
\label{fig1}
\end{center}
\end{figure}

At first, we consider a model space spanned by the five neutron $sdgh$
orbitals placed above doubly closed $^{100}$Sn, so as to compare with
previous realistic shell-model calculations.
The calculated excitation energies of the yrast $2^+$ states
($E^{ex}_{2^+_1}$) and the $B(E2;0^+_1 \rightarrow 2^+_1)$ transition
rates for tin isotopes up to $N=58$ are reported in Fig. \ref{fig1}
and compared with the recent experimental data
\cite{Banu05,Vaman07,Ekstrom08,Guastalla13,Bader13,Doornenbal14}.
It can be seen that, while the experimental behavior of both excitation
energies and $B(E2)$s is reproduced, the observed quadrupole
collectivity is underestimated by our calculations, as evidenced by
the fact that the predicted $E^{ex}_{2^+_1}$s and $B(E2; 0^+_1
\rightarrow 2^+_1)$s are larger and smaller, respectively, than the
experimental ones.
The $B(E2)$s are too small notwithstanding the very large theoretical
neutron effective charges $e_n$, as reported in Table \ref{effchn}.
It should be noted that the theory provides state-dependent effective
charges, and the values relative to the low-lying ($0g_{7/2}$) orbital
exceed the unity, being therefore quite different from the standard
empirical one ($0.5e$).

We stress that this is a result at variance with those obtained in
other regions.
Actually, in our previous works for nuclei above $^{40}$Ca
\cite{Coraggio09c}, $^{48}$Ca \cite{Coraggio14b}, and $^{132}$Sn
\cite{Coraggio09d} cores, starting from the same realistic potential,
we have obtained effective proton and neutron charges close to the
empirical values.
This seems to indicate that for nuclei around $^{100}$Sn relevant
components of the real wavefunction lie outside the chosen model
space, which induces a large renormalization of the theoretical
effective electric-quadrupole operator.

For the sake of completeness, in Fig. \ref{fig1} are also reported the
results obtained with $e_n=0.5e$ (blue squares), which do not 
differ too much from those reported in Ref. \cite{Bader13}, where the
effective interaction has been derived from the chiral N$^3$LO
\cite{Entem03} and NNLO \cite{Ekstrom13} $NN$ potentials.

On the above grounds, we have considered a larger model space so as to
take explicitly into account the $Z=50$ cross-shell excitations of protons
jumping from the $1p_{1/2},0g_{9/2}$ orbitals into the $sdgh$ ones.
Within this large model space we have derived the effective
hamiltonian $H_{\rm eff}^{75}$, where the superscript indicates the
number of proton (seven) and neutron (five) orbitals we have
considered.
In Tables \ref{effchn},\ref{effchp} are reported both theoretical
neutron and proton effective charges, which are closer to the usual
empirical values ($e^{\rm emp}_n =0.5\div 0.8e,~e^{\rm emp}_p=1.5e$).
However, some of the state-dependent $e_n$ are close to unity, which
may be considered an anomalous value with respect to the standard
ones.

As it could be expected, the enlargement of the model space provides
wave functions closer to the real ones, even if the large $e_n$s
indicate that relevant components are still missing.
 
The major difficulty with $H_{\rm eff}^{75}$ is that it cannot be
diagonalized for any tin isotope with up-to-date shell-model codes.
One has then to overcome the computational problem
finding some way to reduce the dimensions of the matrices to be
diagonalized, and consequently make the shell-model calculation
feasible.

This problem has been also faced in Ref. \cite{Banu05}, where the
shell-model calculations have been performed allowing up to $4p-4h$
proton $^{90}$Zr core excitations only.

\begin{table}[H]
\caption{Neutron effective charges of the
  electric quadrupole operator $E2$ for the model space with
  $^{100}$Sn (I) and $^{88}$Sr (II) as cores.}
\begin{ruledtabular}
\begin{tabular}{ccc}
$n_a l_a j_a ~ n_b l_b j_b $ &  $\langle a || e_n || b \rangle $ (I) &  $\langle a || e_n || b \rangle $ (II) \\
\colrule
 $0g_{7/2}~0g_{7/2}$     & 1.20 & 0.94 \\ 
 $0g_{7/2}~1d_{5/2}$     & 1.27 & 0.96 \\ 
 $0g_{7/2}~1d_{3/2}$     & 1.19 & 0.95 \\ 
 $1d_{5/2}~1d_{5/2}$     & 0.81 & 0.94 \\ 
 $1d_{5/2}~1d_{3/2}$     & 0.83 & 0.97 \\ 
 $1d_{5/2}~2s_{1/2}$     & 0.79 & 0.79 \\ 
 $1d_{3/2}~1d_{3/2}$     & 0.87 & 0.96 \\ 
 $1d_{3/2}~2s_{1/2}$     & 0.85 & 0.79 \\ 
 $0h_{11/2}~0h_{11/2}$  & 0.78 & 0.87 \\ 
\end{tabular}
\end{ruledtabular}
\label{effchn}
\end{table}

\begin{table}[H]
\caption{Proton effective charges of the electric quadrupole operator
  $E2$.}
\begin{ruledtabular}
\begin{tabular}{cc}
$n_a l_a j_a ~ n_b l_b j_b $ &  $\langle a || e_p || b \rangle $ \\
\colrule
$0g_{9/2}~0g_{9/2}$     & 1.62 \\ 
 $0g_{9/2}~0g_{7/2}$     & 1.67 \\ 
 $0g_{9/2}~1d_{5/2}$     & 1.60 \\ 
 $0g_{7/2}~0g_{7/2}$     & 1.73\\ 
 $0g_{7/2}~1d_{5/2}$     & 1.74 \\ 
 $0g_{7/2}~1d_{3/2}$     & 1.76 \\ 
 $1d_{5/2}~1d_{5/2}$     & 1.73 \\ 
 $1d_{5/2}~1d_{3/2}$     & 1.72 \\ 
 $1d_{5/2}~2s_{1/2}$     & 1.76 \\ 
 $1d_{3/2}~1d_{3/2}$     & 1.74 \\ 
 $1d_{3/2}~2s_{1/2}$     & 1.76 \\ 
 $0h_{11/2}~0h_{11/2}$  & 1.72 \\ 
\end{tabular}
\end{ruledtabular}
\label{effchp}
\end{table}

\begin{figure}[H]
\begin{center}
\includegraphics[scale=0.38,angle=0]{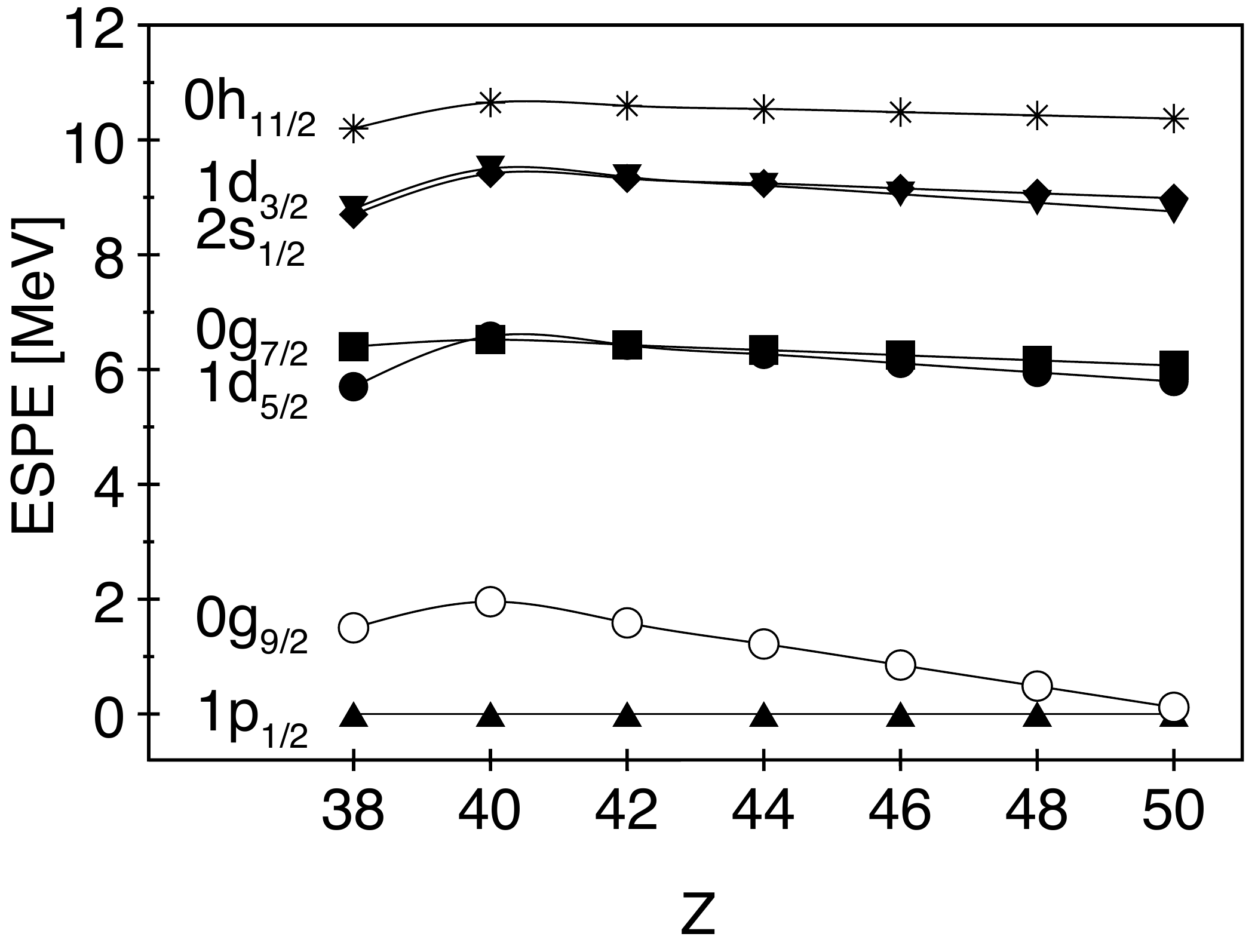}
\caption{Calculated proton effective single-particle energies of
  $H_{\rm eff}^{75}$ as a function of the atomic number $Z$.}
\label{espeZ}
\end{center}
\end{figure}

\begin{figure}[H]
\begin{center}
\includegraphics[scale=0.38,angle=0]{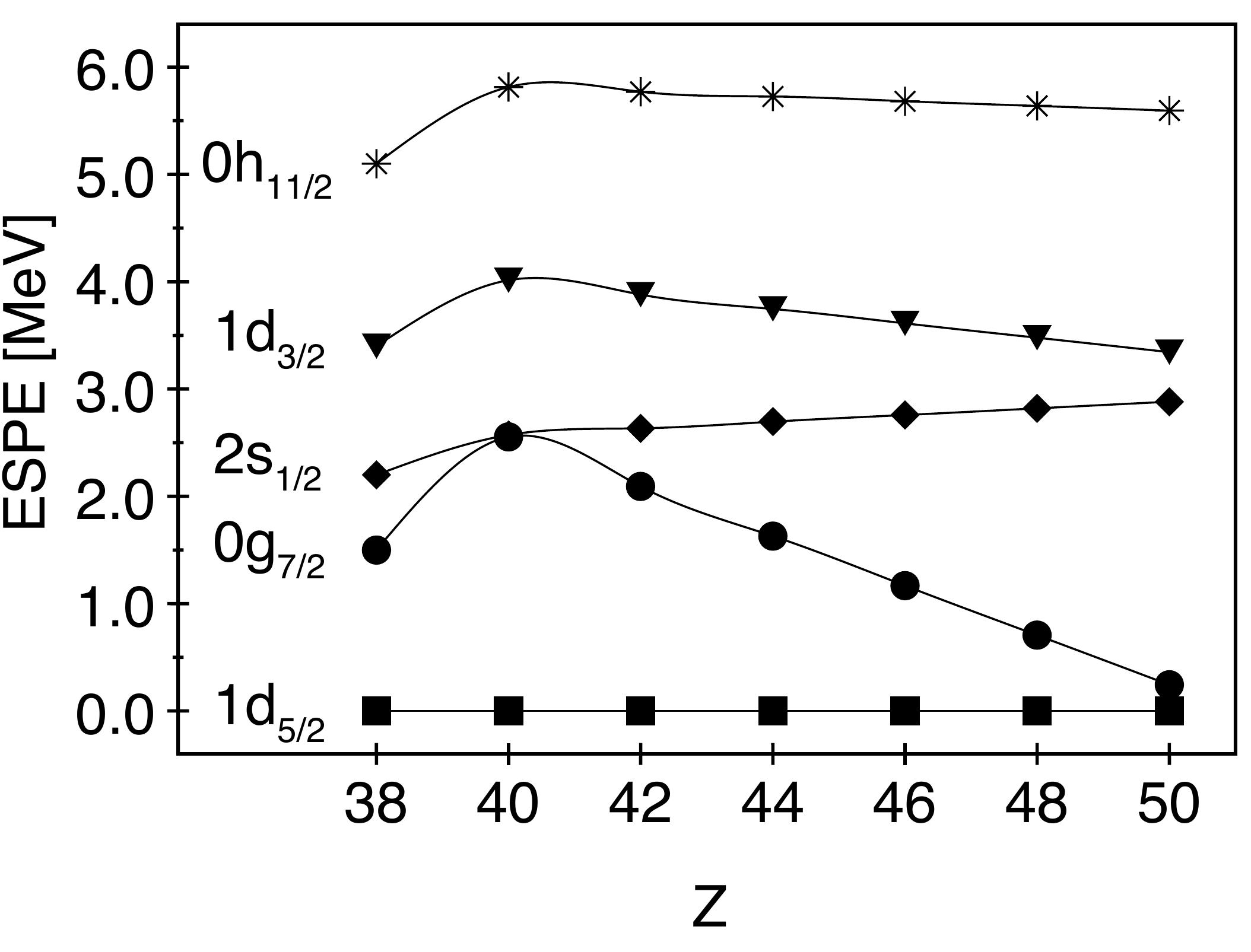}
\caption{Calculated neutron effective single-particle energies of
  $H_{\rm eff}^{75}$ as a function of  $Z$.}
\label{espeN}
\end{center}
\end{figure}

In the present work, we have resorted for the first time to an
approach which, by way of a unitary transformation of $H_{\rm eff}^{75}$,
leads to a new effective hamiltonian defined in a truncated model
space.
The choice of the truncation of the model space is driven by the
behavior, as a function of $Z$ and $N$, of the proton and neutron
effective single-particle energies (ESPE) of the original hamiltonian
$H_{\rm eff}^{75}$, so as to find out what are the most relevant
degrees of freedom to describe the physics of light tin isotopes.
To this end, we report in Figs. \ref{espeZ} and \ref{espeN} the evolution
of both proton and neutron ESPE as a function of $Z$.
From the inspection of Fig. \ref{espeZ}, it can be observed that an
almost constant energy gap provides a separation between the subspace
spanned by the $1p_{1/2},0g_{9/2},1d_{5/2},0g_{7/2}$ proton orbitals
and that spanned by the $2s_{1/2},1d_{3/2},0h_{11/2}$ ones.
This leads to the conclusion that a reasonable truncation is to
consider only the lowest four orbitals, as proton model space.

On the neutron side, Fig. \ref{espeN} evidences that the filling of the
proton $0g_{9/2}$ orbital induces a relevant energy gap at $Z=50$
between the $1d_{5/2},0g_{7/2}$ subspace and that spanned by the
$2s_{1/2},1d_{3/2},0h_{11/2}$ orbitals.
This gap, around 2.4 MeV, traces back to the tensor component of the
proton-neutron interaction that is mainly responsible for the shell
evolution \cite{Otsuka05}.
Our calculated monopole component of the proton-neutron
$0g_{9/2},0g_{7/2}$ interaction is $-0.479$ MeV.

We, therefore, have deemed it reasonable that a neutron model space
spanned only by the $1d_{5/2},0g_{7/2}$ orbitals may provide the
relevant features of the physics of light tin isotopes.

\begin{figure}[ht]
\begin{center}
\includegraphics[scale=0.38,angle=0]{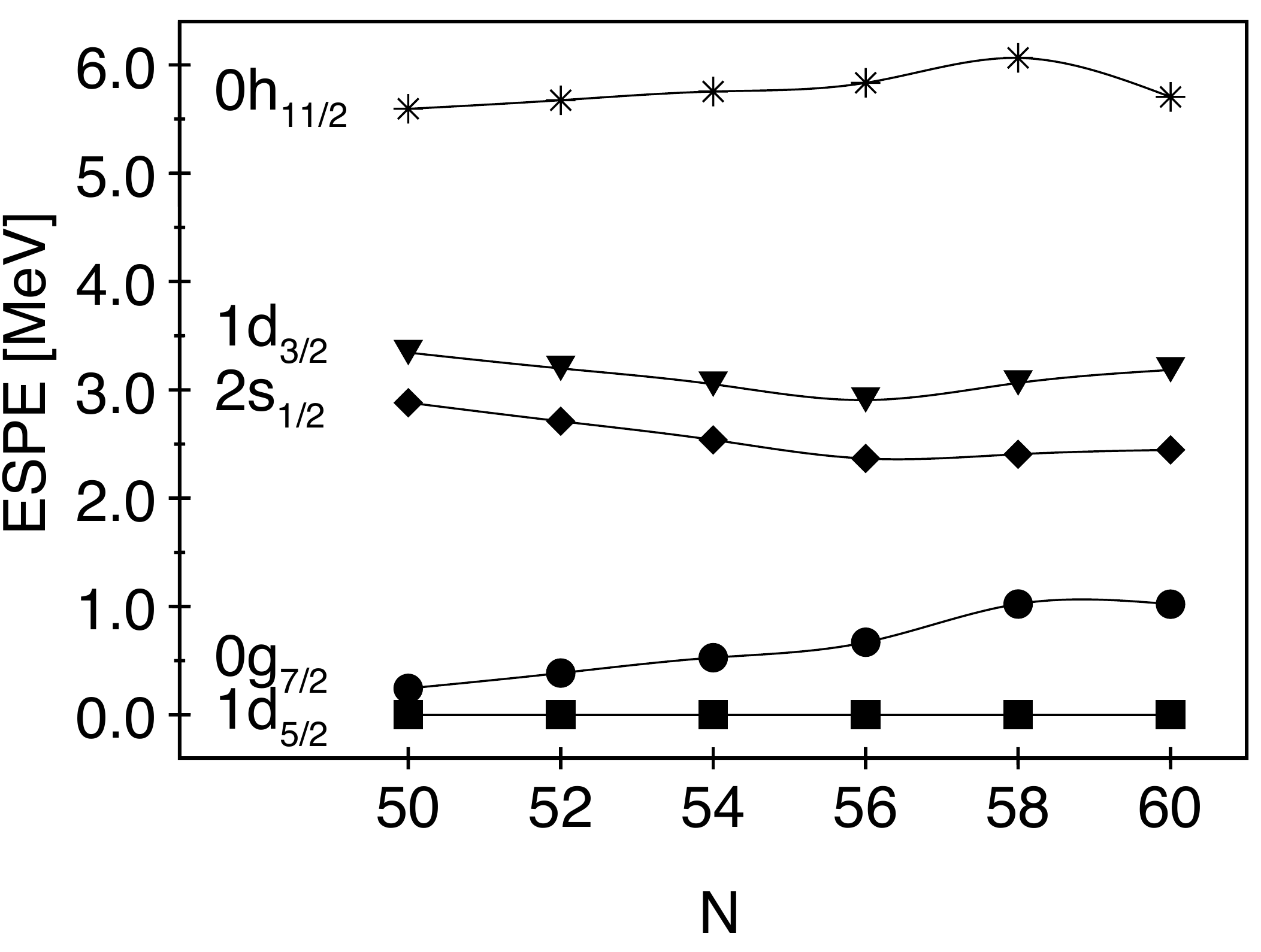}
\caption{Effective single-particle energies of tin isotopes as a
  function of  $N$ calculated with $H_{\rm eff}^{75}$.}
\label{espe_tin}
\end{center}
\end{figure}

Moreover, if we consider the evolution of the neutron ESPE as a
function of $N$ (see Fig. \ref{espe_tin}), it can be observed that
$2s_{1/2},1d_{3/2},0h_{11/2}$ orbitals start to play a more relevant
role from $^{108}$Sn on.
Actually, the reduction of the energy gap between these orbitals and
the $1d_{5/2},0g_{7/2}$ ones, and the progressive filling of the
latter, may reduce the effectiveness of the truncated model space to
describe the physics of tin isotopes above $N=56$.

On these grounds, we have derived a new effective hamiltonian $H_{\rm
  eff}^{42}$, defined within a model space spanned only by the
$1p_{1/2},0g_{9/2},1d_{5/2},0g_{7/2}$ proton and $1d_{5/2},0g_{7/2}$
neutron orbitals, by way of a unitary transformation of $H_{\rm
  eff}^{75}$ (see, for example, Refs. \cite{Suzuki11,Suzuki14}).

It should be pointed out that we have applied this unitary
transformation to the two valence-nucleon systems only, so that the
energy spectra of $^{90}$Zr, $^{90}$Sr, and $^{90}$Y are exactly the
same when diagonalizing $H_{\rm eff}^{75}$ and $H_{\rm eff}^{42}$.
In order to obtain the same outcome for the eigenvalues of $^{102}$Sn,
besides $H_{\rm eff}^{42}$, one should include effective many-body
forces, that can only be obtained by diagonalizing $H_{\rm eff}^{75}$
for this nucleus. 
As already pointed out, this is unfeasible, so we have taken into
account only TBME of $H_{\rm eff}^{42}$.

\begin{figure}[H]
\begin{center}
\includegraphics[scale=0.48,angle=0]{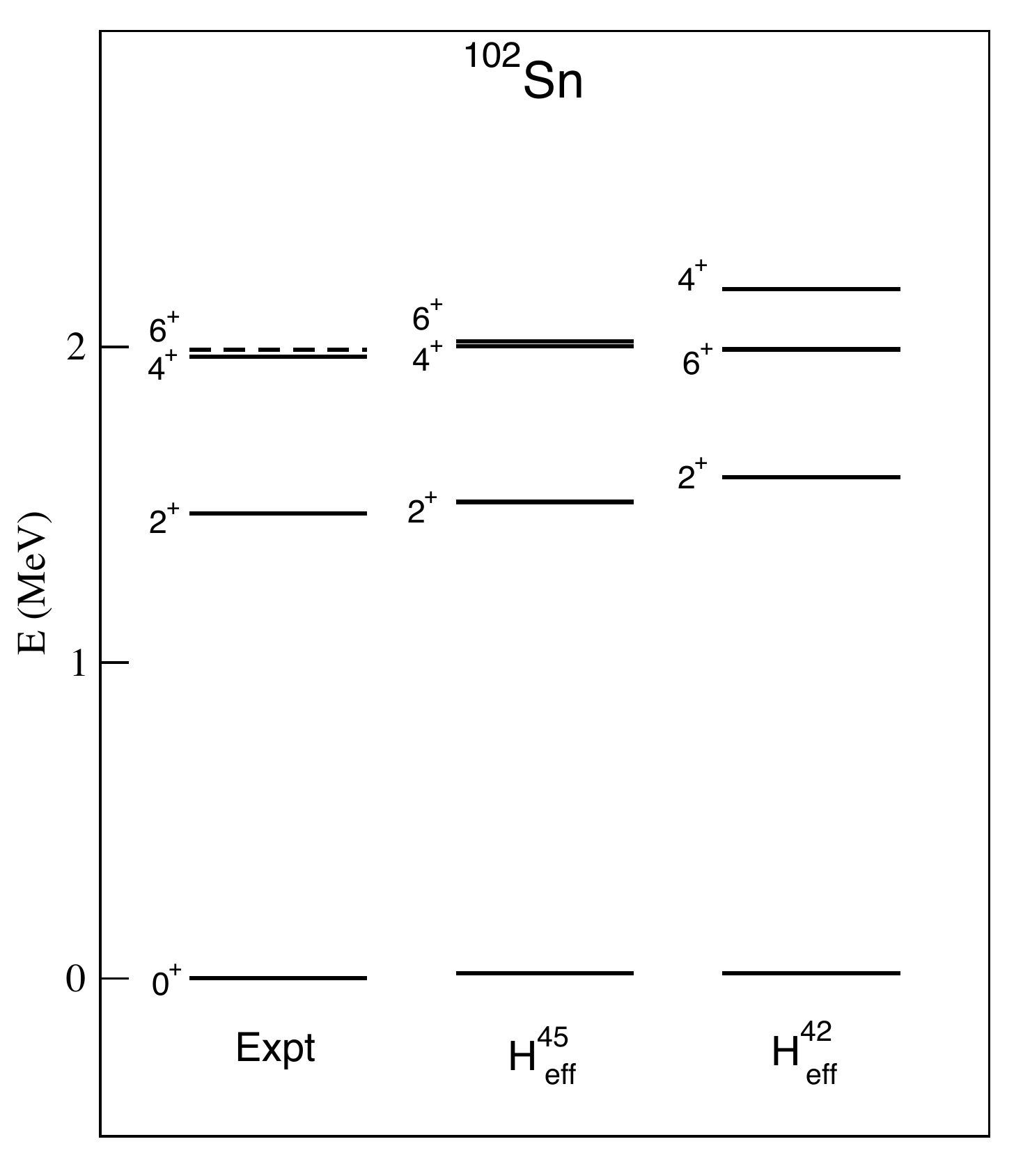}
\caption{Experimental \cite{ensdf} and theoretical spectra of
  $^{102}$Sn calculated with $H_{\rm eff}^{45}$ and $H_{\rm
    eff}^{42}$ (see text for details).}
\label{102Sn}
\end{center}
\end{figure}

It is obvious that the larger is the chosen subspace the smaller is
the role of these effective many-body components.
First of all, it may be worth verifying the reliability of our
truncation scheme for tin isotopes.
To this end we have derived another hamiltonian, starting from $H_{\rm
  eff}^{75}$, defined in the full $sdgh$ neutron model space but with
the same proton model space of $H_{\rm eff}^{42}$.
This is the largest model space in which we can manage to diagonalize
the shell-model hamiltonian of $^{102}$Sn.
We dub this effective hamiltonian $H_{\rm eff}^{45}$, and in
Fig. \ref{102Sn} we compare the low-energy spectra of $^{102}$Sn obtained
by means of $H_{\rm eff}^{45}$ and $H_{\rm eff}^{42}$ with the
observed one \cite{ensdf}.

It can be noted that  $H_{\rm eff}^{45}$ is able to reproduce quite
well the experimental spectrum, and that the spectrum calculated with
$H_{\rm eff}^{42}$ is in a good agreement with results obtained with
$H_{\rm eff}^{45}$.
It is very relevant, for the subject of our study, to point out that
the $B(E2;0^+_1 \rightarrow 2^+_1)$ calculated with $H_{\rm
  eff}^{42}$ is $1065~e^2{\rm fm}^4$, very close to the value of
$1135~e^2{\rm fm}^4$ obtained with $H_{\rm eff}^{45}$.
This supports the adequacy of our truncation scheme when using
$H_{\rm eff}^{42}$  and increasing the number of valence neutrons. 
This is an important result since our calculations, performed by way
of the Oslo shell-model code \cite{EngelandSMC},  cannot be extended
to heavier tin isotopes using $H_{\rm eff}^{45}$.

\begin{figure}[H]
\begin{center}
\includegraphics[scale=0.38,angle=0]{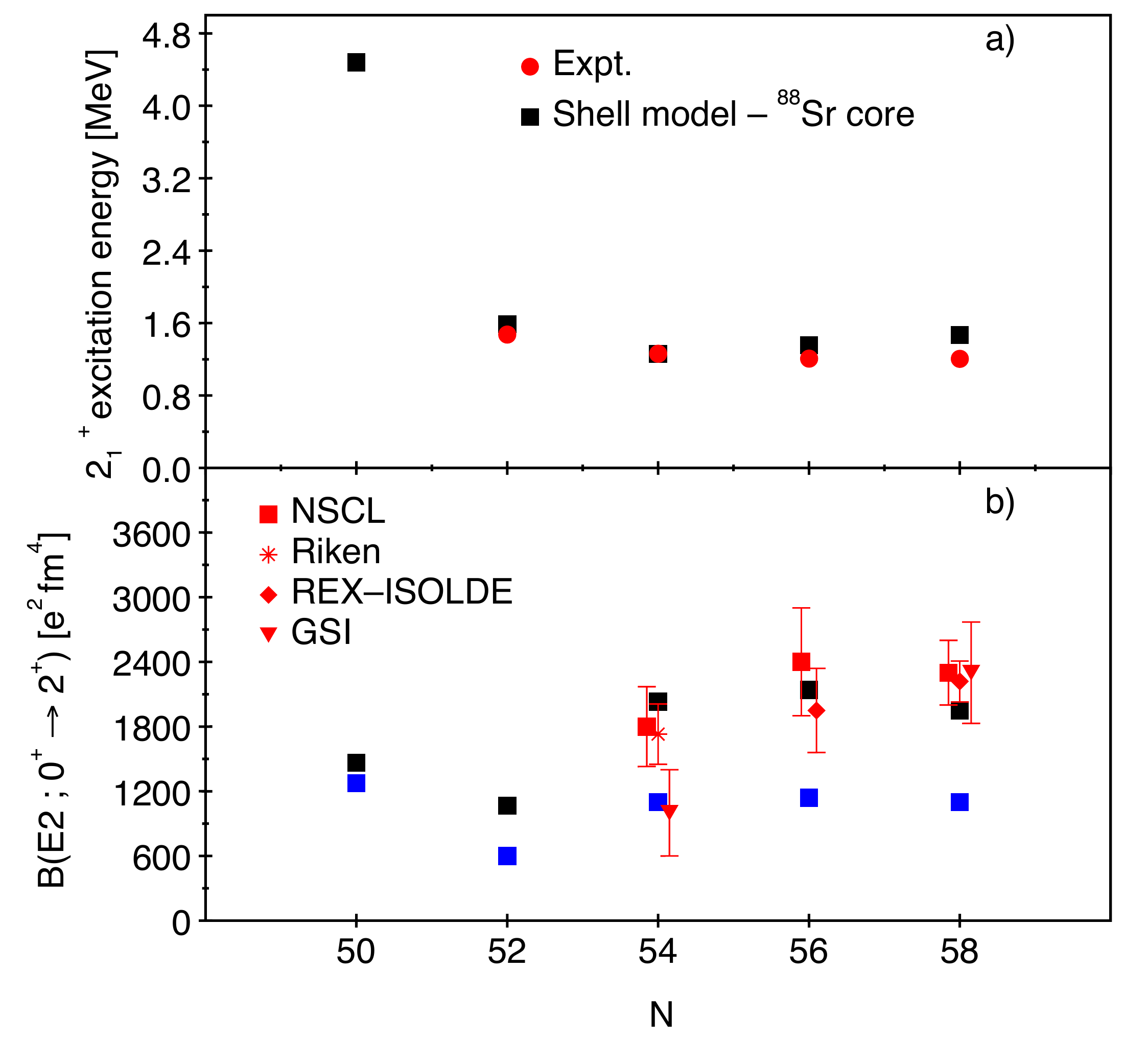}
\caption{(Color online) Same as in Fig. \ref{fig1}, but with
  shell-model results obtained with $H_{\rm eff}^{42}$. Blue squares
  refer to calculations with $e^{emp}_n=0.5,~e^{emp}_p=1.5$.}
\label{results}
\end{center}
\end{figure}

In Fig. \ref{results} we report the experimental $E^{ex}_{2^+_1}$ and
$B(E2;0^+_1 \rightarrow 2^+_1)$ (red symbols) for the tin isotopes up to
$N=58$, and compare them with the results obtained using $H_{\rm
  eff}^{42}$ (black squares).
We have also included our prediction for the closed shell $^{100}$Sn,
where only proton degrees of freedom are taken into account.
As in Fig. \ref{fig1}, we have included the calculated $B(E2)$s when
using $e^{\rm emp}_n =0.5e,~e^{\rm emp}_p=1.5e$.

We see that the shell-model calculations with $H_{\rm eff}^{42}$ are
able to reproduce quite well the experimental $E^{ex}_{2^+_1}$ and,
when employing the theoretical effective charges, the $B(E2;0_1^+
\rightarrow 2^+_1)$ up to $A=106$ and, consequently, the onset of
collectivity from $^{102}$Sn on, driven by the $Z=50$ cross-shell
excitations.
As a matter of fact, the wavefunctions evidence a depletion of the
proton $0g_{9/2}$ orbital from $^{102}$Sn to $^{106}$Sn, as testified
by the occupation numbers reported in Table \ref{occupations}.

It is worth to point out that the agreement with experiment for
$^{108}$Sn deteriorates with respect to lighter isotopes, owing to the
the fact that, as already mentioned, the influence of neutron
$1d_{3/2},2s_{1/2},0h_{11/2}$ orbitals starts to play a non-negligible
role.

\begin{table}[H]
\caption{Occupation numbers of proton
  $1p_{1/2},0g_{9/2},0g_{7/2},1d_{5/2}$ of $^{102-108}$Sn $J=0^+_1, 2^+_1$ state, calculated
  with model space III (see text for details).}
\label{occupations}
\begin{ruledtabular}
\begin{tabular}{ccccc}
\hline
 Orbital & $^{102}$Sn & $^{104}$Sn & $^{106}$Sn & $^{108}$Sn \\
\hline
~&~& $J=0^+$ & ~& ~\\
\hline
 $\pi 1p_{1/2}$ & 1.97 & 1.98 & 1.98 & 1.98 \\
 $\pi 0g_{9/2}$ & 9.55 & 9.38 &  9.36 &  9.38 \\
 $\pi 0g_{7/2}$ & 0.26 & 0.28 &  0.27 &  0.25 \\
 $\pi 1d_{5/2}$ & 0.22 & 0.37 &  0.40 &  0.39 \\
\hline
~&~& $J=2^+$ & ~& ~\\
\hline
 $\pi 1p_{1/2}$ & 1.98 & 1.98 & 1.98 & 1.98 \\
 $\pi 0g_{9/2}$ & 9.50 & 9.23 &  9.21 &  9.27 \\
 $\pi 0g_{7/2}$ & 0.25 & 0.28 &  0.27 &  0.24 \\
 $\pi 1d_{5/2}$ & 0.28 & 0.51 &  0.54 &  0.51 \\
\hline
\end{tabular}
\end{ruledtabular}
\end{table}

In summary, we have performed a shell-model study of light tin
isotopes starting from a realistic $NN$ potential, where all results
have been obtained without resorting to any empirical parameter.
The main features of present work may be itemized as follows:
\begin{itemize}
\item We have confirmed the crucial role of the $Z=50$
  cross-shell excitations to obtain a satisfactory description of tin
  isotopes. This implies the use of a large shell-model space
  including both proton and neutron orbitals.
\item We have followed an original double-step approach to reduce the
  computational complexity of the shell-model problem. This is based
  on the study of the ESPE of the large-scale hamiltonian, so as to
  identify the most relevant degrees of freedom to be taken into
  account in the construction of a truncated shell-model
  hamiltonian. To this end, a unitary transformation is employed.
\item We have highlighted the role of theoretical effective
  charges in reproducing the quadrupole collectivity of the $B(E2)$s.
\item We have presented some predictions for $^{100,102}$Sn
  spectroscopic properties, which may provide guidance for future
  experiments. 
\end{itemize}

\bibliographystyle{apsrev}
\bibliography{biblio}

\end{document}